\documentclass[%
 reprint,
 amsmath,amssymb,
 aps,
]{revtex4-1}

\usepackage{graphicx}
\usepackage{dcolumn}
\usepackage{bm}
\usepackage{color}

\begin{document}

\preprint{APS/123-QED}

\title{Nonlinear mode coupling in whispering-gallery-mode resonators}

\author{Giuseppe D'Aguanno}
\email{giusdag@umbc.edu}
\author{Curtis R. Menyuk}%
 \email{menyuk@umbc.edu}
\affiliation{%
 Department of Computer Science and Electrical Engineering, University of Maryland, 1000 Hilltop Circle, Baltimore, Maryland
21250, United States}%

%
%
%

\begin{abstract}
We present a first principle derivation of the coupled nonlinear Schr\"{o}dinger equations that govern  the  interaction between two families of  modes with different transverse profiles in a generic  whispering-gallery-mode resonator. We find regions of modulational instability and the existence of trains of bright solitons both in the normal and in the anomalous dispersion regime. 
\begin{description}
\item[PACS numbers]
42.62.Eh, 42.65.Hw, 42.65.Sf, 42.65.Tg 
\end{description}
\end{abstract}

\maketitle


\section{Introduction}

\indent Whispering gallery modes (WGMs) have been a source of fascination to physicists since at least the work of Lord Rayleigh in 1910 \cite{Rayleigh}, when he explained the phenomenon that a whisper in one end of the gallery of St.~Paul's Cathedral could be heard at the other end. This phenomenon has practical implications as well, since WGMs in resonators with cylindrical or spherical symmetry can have very high quality (Q) factors. In the last two decades, micro-cavity resonators have been increasingly  used to generate and filter narrowband light \cite {Vahala,Notomi}. The quest to obtain pure optical frequency sources was revolutionized in 2000 with the invention of locked frequency comb lasers \cite {Ranka,Diddams}. This revolution was enabled by technology that allowed the inventors of the comb laser to achieve a factor of two (an octave) of bandwidth. However, these sources have the drawback that they are typically  bulky and expensive. Today, we may be on the verge of a second revolution in frequency generation. In fact, in the past eight years it has been demonstrated that it is possible to use WGM micro-resonators to generate solitons and hence broadband combs \cite {DelHaye,Herr} and, within the past year, they have been used to generate nearly an octave of bandwidth and to lock the carrier-envelope phase \cite {Yi}. However, the process by which these modes are generated remains poorly understood. It has been demonstrated that single modes are governed by the Lugiato-Lefever equation (LLE) \cite {Matsko, Coen, Chembo}; however, there is recent experimental evidence that mode coupling can play a critical role in obtaining a broad bandwidth comb \cite {Weiner,Gaeta}. That is particularly the case at optical and near-ultraviolet (UV) wavelengths at which the modes typically have normal dispersion and solitons cannot be obtained from a single mode.\\
\indent In this Letter, we present for the first time coupled LLEs that  describe the coupling of two WGMs in a resonator in which chromatic dispersion, the Kerr nonlinearity, and an external pump are all present. These equations resemble the coupled nonlinear Schr\"{o}dinger equations (NLSEs) that describe mode coupling in optical fibers and waveguides \cite {Berkhoer} and Bose-Einstein condensates \cite {Ma}. However, the presence of the pump and the periodicity of the resonator change the equations in a fundamental way. It is no longer possible to remove the phase velocity difference from the equations by separately shifting the central frequencies of each of the modes because that changes their frequency difference from the pump.  It is natural to choose the central frequency for each mode so that it matches the pump frequency. Moreover, periodic boundary conditions must be imposed along the spatial coordinate of the resonator. Among consequences, we will find that solitons typically form on a broad pedestal, and it is possible to obtain the modulational instability (MI)  in the normal dispersion regime. While our own focus is on applications to micro-resonators, we note that the LLE has a broad range of applications throughout physics. Since nonlinear mode coupling is a phenomenon that occurs in many physical systems, we anticipate that the equations and the phenomena that we describe here will have a similarly broad range of applications.
\section{Results and Discussion}
\indent Our aim in this paper is twofold: First, we provide an \textit{ab initio}
derivation of the set of externally driven, coupled, damped, NLSEs, or coupled LLEs, that account for nonlinear mode coupling in a WGM resonator with a Kerr  ($\chi^{(3)})$ nonlinearity. Second, we
apply the equations that we derive to discuss the onset of the MI and the formation of   trains of mode-locked bright solitons.\\
\indent The two coupled LLEs are given by  
\begin {equation}\normalsize
\frac{\partial \Psi^{(j)}}{\partial t}=\sum\limits_{k=1}^{n}\ (-i)^{k+1}\frac{\zeta_k^{(j)}}{k!}\frac{\partial^k \Psi^{(j)}}{\partial \theta^k}-\frac{1}{\tau^{(j)}}\Psi^{(j)}(\theta,t)
\notag
\end {equation}
\vspace{-6mm}
\begin {equation}
+i\frac{P_{\bar m}^{(j)}\omega_p}{2} \exp[{i\delta\omega^{(j)}t}]\hspace{1.5cm}
 \notag
 \end {equation}
\vspace{-6mm}
\begin {equation}
+i\chi^{(3)} \omega_{\bar m}^{(j)}\Psi^{(j)}\sum\limits_{l=1}^{2}D^{(j,l)}|\Psi^{(l)}|^2 \:,
 \tag{1}
\end {equation}
with $j=1,2$, and $n$ an integer greater than 1, where
\begin {equation}
D^{(j,l)}=\frac{2^{(|j-l|-1)}}{\epsilon_{r}(\omega_{\bar m}^{(j)})V_c}\times\hspace{5.5cm}
\notag 
\end {equation}
\vspace{-6mm}
\begin {equation}
\hspace{1cm}\int \limits_{V_c}[|\textbf{F}_{\bar m}^{(j)}|^2|\textbf{F}_{\bar m}^{(l)}|^2+
|\textbf{F}_{\bar m}^{(j)}\cdot\textbf{F}_{\bar m}^{(l)}|^2+|\textbf{F}_{\bar m}^{(j)}\cdot\textbf{F}_{\bar m}^{(l)*}|^2]\,dV,
 \tag{2}
 \end {equation}
are the overlap integrals of the two dominant modes, $V_c$ is the resonator volume, $\bar m$ and $\Psi^{(j)} (\theta,t)$ are respectively the azimuthal number and the spatio-temporal envelope of the dominant modes, $\theta$ is the azimuthal coordinate of the WGM resonator, $\tau^{(j)}$ is the  photon lifetime in the cavity, $\delta\omega^{(j)}=\omega_{\bar m}^{(j)}-\omega_p$ is the detuning of the frequency of the pump field with respect to the frequency of the dominant mode, $\omega_{\bar m}^{(j)}$ and $\omega_p$ are respectively the dominant mode frequency and the pump frequency, $P_{\bar m}^{(j)}$ is the pump field coupled to the resonator, $\chi^{(3)}>0$ is the self-focusing Kerr nonlinearity of the resonator material, and $\zeta_1^{(j)}=(1/2) (\omega_{\bar m+1}^{(j)}-\omega_{\bar m-1}^{(j)})$ is  the free spectral range (FSR) of the resonator calculated at $\omega_{\bar m}^{(j)}$. The coefficient $\zeta_2^{(j)}=\omega_{\bar m+1}^{(j)}-2\omega_{\bar m}^{(j)}+\omega_{\bar m-1}^{(j)}$ denotes, at lowest order, the deviation from  equidistance of the eigenfrequencies adjacent to  $\omega_{\bar m}^{(j)}$ and plays a role analogous to the group velocity dispersion (GVD) of a standard optical fiber, while $\zeta_k^{(j)}$ with $k>2$ are higher-order dispersion coefficients.  We note that $\zeta_2^{(j)}<0$ corresponds to the normal dispersion regime in which the group velocity decreases for increasing frequencies, while $\zeta_2^{(j)}>0$ corresponds to anomalous dispersion. We also introduce, as is customary, the GVD parameter $\beta_2^{(j)}=-\zeta_2^{(j)}=2\omega_{\bar m}^{(j)}-\omega_{\bar m+1}^{(j)}-\omega_{\bar m-1}^{(j)}$, so that the dispersion is normal when $\beta_2^{(j)}>0$ and anomalous when $\beta_2^{(j)}<0$.  A detailed derivation of Eq.~(1) and (2) from Maxwell's equations is given in the Appendix.
 We introduce the following dimensionless variables and parameters: $\tau=t/\bar\tau$, the time normalized to the average cavity photon lifetime of the two modes $[\bar\tau=(\tau^{(1)}+\tau^{(2)})/2]$; $\alpha^{(j)}=\delta\omega^{(j)}\bar\tau$, the normalized detuning; $\bar\zeta_k^{(j)}=\zeta_k^{(j)}\bar\tau$, the normalized dispersion coefficients;  $Q\cong\bar\tau\omega_{\bar m}^{(j)}/2\cong\bar\tau\omega_p/2$ the cavity Q-factor referred to the average cavity photon lifetime; $\psi^{(j)}=\sqrt{2\chi^{(3)}Q}e^{-i\alpha^{(j)\tau}}\Psi^{(j)}$, the dimensionless field envelope; and  $h^{(j)}={P_{\bar m}^{(j)}}\sqrt{2\chi^{(3)}Q^3}$ the dimensionless pump field. Using the coordinate transformation $\theta\rightarrow\theta-\bar\zeta_1^{(\rm{av})}\tau    \:\rm{mod}[2\pi]$, where $\bar\zeta_1^{(\rm{av})}=(\bar\zeta_1^{(1)}+\bar\zeta_1^{(2)})/2$ is the average group velocity, and only keeping terms up to $n=2$, we find that Eq.~(1) becomes
 \begin {equation}\normalsize
\frac{\partial \psi^{(j)}}{\partial \tau}=\delta^{(j)}\frac{\partial \psi^{(j)}}{\partial \theta}-i\frac{\bar\beta_2^{(j)}}{2}\frac{\partial^2 \psi^{(j)}}{\partial \theta^2}-\left(\frac{\bar\tau}{\tau^{(j)}}+i\alpha^{(j)}\right)\psi^{(j)}
\notag
\end {equation}
\vspace{-6mm}
\begin {equation}\normalsize
+ih^{(j)}+i\psi^{(j)}\sum\limits_{k=1}^{2}D^{(j,k)}|\psi^{(k)}|^2,\hspace{11mm}
\tag{3}
 \end {equation}
 with $j=1,2$. The quantity $\delta^{(j)}=\bar\zeta_1^{(\rm{av})}-\bar\zeta_1^{(j)}$ is the group velocity mismatch (GVM) of the two envelope fields with respect to the average group velocity and $\bar\beta_2^{(j)}=-\bar\zeta_2^{(j)}$ is the GVD parameter. Note that in the retarded coordinate system traveling at the average GVD  we have $\delta^{(1)}=-\delta^{(2)}=\delta$.  Equation (3) has the form of two coupled LLEs \cite {Lugiato,Chembo}, plus the additional term $\delta^{(j)}\partial \psi^{(j)}/\partial \theta$ representing the GVM. 
\indent We now discuss the particular situation of degenerate interacting modes. In this case, we find $\delta=0$, $\bar\beta_2^{(1)}=\bar\beta_2^{(2)}=\bar\beta_2$, $\bar\tau/\tau^{(1)}=\bar\tau/\tau^{(2)}=1$ and $\alpha^{(1)}=\alpha^{(2)}=\alpha$. It is useful to rescale Eq.~(3) by introducing the following variables and parameters: $\sigma=\theta/{(|\bar\beta_{2}|)^{1/2}}$, the scaled azimuthal coordinate, $U^{(j)}=\sqrt{D^{(1,1)}}\psi^{(j)}$, the scaled field envelope, $P^{(j)}=D^{(1,1)}h^{(j)2}$, the scaled pump power, $g_{\rm c}=D^{(2,1)}/D^{(1,1)}$, the cross-coupling parameter, and $g_{\rm s}=D^{(2,2)}/D^{(1,1)}$, the self-coupling parameter. Equation (3) now becomes
 \begin {equation}\normalsize
i\frac{\partial U^{(1)}}{\partial \tau}-\frac{\rm{sgn}(\bar\beta_{2})}{2}\frac{\partial^2 U^{(1)}}{\partial \sigma^2}+(i-\alpha) U^{(1)}
\notag
\end {equation}
\vspace{-6mm}
\begin {equation}
+U^{(1)}(|U^{(1)}|^{2}+g_{\rm c}|U^{(2)}|^{2})=-\sqrt{P^{(1)}}\:,
 \tag{4.a}
  \end {equation}
\begin {equation}\normalsize
i\frac{\partial U^{(2)}}{\partial \tau}-\frac{\rm{sgn}(\bar\beta_{2})}{2}\frac{\partial^2 U^{(2)}}{\partial \sigma^2}+(i-\alpha) U^{(2)}
\notag
\end {equation}
\vspace{-6mm}
\begin {equation}
+U^{(2)}(g_{\rm c}|U^{(1)}|^{2}+g_{\rm s}|U^{(2)}|^{2})=-\sqrt{P^{(2)}}\:,
 \tag{4.b}
  \end {equation}
where $\rm{sgn}(\cdot)$ is the sign function. The reader will note a formal analogy between our Eq.~(4) and the equations that describe self-focusing of waves with different polarizations in a Kerr medium \cite {Berkhoer}. We also note that in the limit of negligible cross-coupling, i.e. $g_{\rm c}\rightarrow 0 $, Eq.~(4) decomposes into two uncoupled LLEs, as expected. Since mode coupling has already been observed to play an important role in microresonators \cite {Weiner,Gaeta}, we anticipate that these equations and their extension to more than two modes will have a broad range of applications. Here,  we focus on the  MI for two particular classes of continuous-wave (CW), spatially homogeneous solutions  admitted by Eq.~(4). We write these solutions as $U_{0}^{(j)}$. The formation of a periodic pulse train is generally initiated by the MI of the CW solutions \cite {Zakharov,Hasegawa}. As is usually done when the MI is studied, we first write the field envelope as $U^{(j)}=[U_{0}^{(j)}+v^{(j)}+iw^{(j)}]$, where $v^{(j)}(\sigma,\tau)$ and $w^{(j)}(\sigma,\tau)$ are small perturbations, and we next linearize Eq.~(4) around $U_{0}^{(j)}$ to obtain:
 \begin {equation}\normalsize
-\frac{\partial w^{(1)}}{\partial \tau}-\frac{\rm{sgn}(\bar\beta_{2})}{2}\frac{\partial^2 v^{(1)}}{\partial \sigma^2}+[2|U_{0}^{(1)}|^2+g_{\rm{c}}|U_{0}^{(2)}|^2-\alpha
\notag
\end {equation}
\vspace{-6mm}
\begin {equation}
+{\rm{Re}}(U_0^{(1)2})] v^{(1)}+w^{(1)}[{\rm{Im}}(U_0^{(1)2})-1]
 \notag
  \end {equation}
\vspace{-6mm}
\begin {equation}
+2g_{\rm{c}}[{\rm{Re}}(U_0^{(1)}) {\rm{Re}}(U_0^{(2)})v^{(2)}+{\rm{Re}}(U_0^{(1)}) {\rm{Im}}(U_0^{(2)})w^{(2)}]=0\:,
 \tag {5.a}
  \end {equation}
\begin {equation}\normalsize
\frac{\partial v^{(1)}}{\partial \tau}-\frac{\rm{sgn}(\bar\beta_{2})}{2}\frac{\partial^2 w^{(1)}}{\partial \sigma^2}+[2|U_{0}^{(1)}|^2+g_{\rm{c}}|U_{0}^{(2)}|^2-\alpha
\notag
\end {equation}
\vspace{-6mm}
\begin {equation}
-{\rm{Re}}(U_0^{(1)2})] w^{(1)}+v^{(1)}[{\rm{Im}}(U_0^{(1)2})+1]
 \notag
  \end {equation}
\vspace{-6mm}
\begin {equation}
+2g_{\rm{c}}[{\rm{Im}}(U_0^{(1)}) {\rm{Re}}(U_0^{(2)})v^{(2)}+{\rm{Im}}(U_0^{(1)}) {\rm{Im}}(U_0^{(2)})w^{(2)}]=0\:,
 \tag {5.b}
\end{equation}
\begin {equation}\normalsize
-\frac{\partial w^{(2)}}{\partial \tau}-\frac{\rm{sgn}(\bar\beta_{2})}{2}\frac{\partial^2 v^{(2)}}{\partial \sigma^2}+[g_{\rm{c}}|U_{0}^{(1)}|^2+2g_{\rm{s}}|U_{0}^{(2)}|^2-\alpha
\notag
\end {equation}
\vspace{-6mm}
\begin {equation}
+g_{\rm{s}}{\rm{Re}}(U_0^{(2)2})] v^{(2)}+w^{(2)}[g_{\rm{s}}{\rm{Im}}(U_0^{(2)2})-1]
 \notag
  \end {equation}
\vspace{-6mm}
\begin {equation}
+2g_{\rm{c}}[{\rm{Re}}(U_0^{(1)}) {\rm{Re}}(U_0^{(2)})v^{(1)}+{\rm{Im}}(U_0^{(1)}) {\rm{Re}}(U_0^{(2)})w^{(1)}]=0\:,
 \tag {5.c}
\end {equation}
\begin {equation}\normalsize
\frac{\partial v^{(2)}}{\partial \tau}-\frac{\rm{sgn}(\bar\beta_{2})}{2}\frac{\partial^2 w^{(2)}}{\partial \sigma^2}+[g_{\rm{c}}|U_{0}^{(1)}|^2+2g_{\rm{s}}|U_{0}^{(2)}|^2-\alpha
\notag
\end {equation}
\vspace{-6mm}
\begin {equation}
-g_{\rm{s}}{\rm{Re}}(U_0^{(2)2})] w^{(2)}+v^{(2)}[g_{\rm{s}}{\rm{Im}}(U_0^{(2)2})+1]
 \notag
  \end {equation}
\vspace{-6mm}
\begin {equation}
+2g_{\rm{c}}[{\rm{Re}}(U_0^{(1)}) {\rm{Im}}(U_0^{(2)})v^{(1)}+{\rm{Im}}(U_0^{(1)}) {\rm{Im}}(U_0^{(2)})w^{(1)}]=0\:.
 \tag {5.d}
\end {equation}

We look for traveling-wave solutions in the form $v^{(j)}={\rm{Re}}\{x^{(j)}\exp[i(K \sigma-\Omega \tau)]\}$ and $w^{(j)}={\rm{Re}}\{y^{(j)}\exp[i(K \sigma-\Omega \tau)]\}$, where $\Omega$ is the frequency shift with respect to the carrier frequency of the dominant mode and $K$ is the corresponding shift in the wavenumber. Substituting the traveling wave solutions into Eq.~(5), we arrive at the following system of linear, homogeneous, algebraic equations:
\begin {equation}\normalsize
A[x^{(1)},y^{(1)},x^{(2)},y^{(2)}]^T=0\:,
\tag{6}
\end {equation}
where $A$ is a $4 \times 4$ matrix whose elements $a_{j,k}$ are given by:
\begin {equation}\normalsize
a_{1,1}=\frac{K^2\rm{sgn}(\bar\beta_{2})}{2}+2|U_{0}^{(1)}|^2+g_{\rm{c}}|U_{0}^{(2)}|^2-\alpha+{\rm{Re}}(U_0^{(1)2})\:,
\tag{7.a}
\end {equation}
\begin {equation}\normalsize
a_{1,2}=i\Omega+{\rm{Im}}(U_0^{(1)2})-1\:,
\tag{7.b}
\end {equation}
\begin {equation}\normalsize
a_{1,3}=2g_{\rm{c}}{\rm{Re}}(U_0^{(1)}) {\rm{Re}}(U_0^{(2)})\:,
\tag{7.c}
\end {equation}
\begin {equation}\normalsize
a_{1,4}=2g_{\rm{c}}{\rm{Re}}(U_0^{(1)}) {\rm{Im}}(U_0^{(2)})\:,
\tag{7.d}
\end {equation}
\begin {equation}\normalsize
a_{2,1}=-i\Omega+{\rm{Im}}(U_0^{(1)2})+1\:,
\tag{7.e}
\end {equation}
\begin {equation}\normalsize
a_{2,2}=\frac{K^2\rm{sgn}(\bar\beta_{2})}{2}+2|U_{0}^{(1)}|^2+g_{\rm{c}}|U_{0}^{(2)}|^2-\alpha-{\rm{Re}}(U_0^{(1)2})\:,
\tag{7.f}
\end {equation}
\begin {equation}\normalsize
a_{2,3}=2g_{\rm{c}}{\rm{Im}}(U_0^{(1)}) {\rm{Re}}(U_0^{(2)})\:,
\tag{7.g}
\end {equation}
\begin {equation}\normalsize
a_{2,4}=2g_{\rm{c}}{\rm{Im}}(U_0^{(1)}) {\rm{Im}}(U_0^{(2)})\:,
\tag{7.h}
\end {equation}
\begin {equation}\normalsize
a_{3,1}=2g_{\rm{c}}{\rm{Re}}(U_0^{(1)}) {\rm{Re}}(U_0^{(2)})\:,
\tag{7.i}
\end {equation}
\begin {equation}\normalsize
a_{3,2}=2g_{\rm{c}}{\rm{Im}}(U_0^{(1)}) {\rm{Re}}(U_0^{(2)})\:,
\tag{7.j}
\end {equation}
\begin {equation}\normalsize
a_{3,3}=\frac{K^2\rm{sgn}(\bar\beta_{2})}{2}+g_{\rm{c}}|U_{0}^{(1)}|^2+2g_{\rm{s}}|U_{0}^{(2)}|^2-\alpha+g_{\rm{s}}{\rm{Re}}(U_0^{(2)2})
\tag{7.k}
\end {equation}
\begin {equation}\normalsize
a_{3,4}=i\Omega+g_{\rm{s}}{\rm{Im}}(U_0^{(2)2})-1\:,
\tag{7.l}
\end {equation}
\begin {equation}\normalsize
a_{4,1}=2g_{\rm{c}}{\rm{Re}}(U_0^{(1)}) {\rm{Im}}(U_0^{(2)})\:,
\tag{7.m}
\end {equation}
\begin {equation}\normalsize
a_{4,2}=2g_{\rm{c}}{\rm{Im}}(U_0^{(1)}) {\rm{Im}}(U_0^{(2)})\:,
\tag{7.n}
\end {equation}
\begin {equation}\normalsize
a_{4,3}=-i\Omega+g_{\rm{s}}{\rm{Im}}(U_0^{(2)2})+1\:,
\tag{7.o}
\end {equation}
\begin {equation}\normalsize
a_{4,4}=\frac{K^2\rm{sgn}(\bar\beta_{2})}{2}+g_{\rm{c}}|U_{0}^{(1)}|^2+2g_{\rm{s}}|U_{0}^{(2)}|^2-\alpha-g_{\rm{s}}{\rm{Re}}(U_0^{(2)2})\:.
\tag{7.p}
\end {equation}
 The compatibility condition for the existence of traveling-wave solutions,{\it{i.e.}} $\det(A)=0$,  yields  the dispersion relation, $\Omega(K)$. The MI occurs at those values of $K$ at which  ${\rm{Re}}{(i\Omega)}<0$, which physically give rise to an exponential growth of the amplitude of the traveling waves. Due to the $2\pi$-periodicity of the system in the azimuthal coordinate $\theta$, the wavenumber $K$ can only assume discrete values, $K=p{(|\bar\beta_2|)^{1/2}}$, where $p=m-\bar m=\pm1,\pm2,...$ is the shift of the mode number of the perturbation with respect to the mode number $\bar m$ of the dominant mode.\\
\indent We find by substitution into Eq.~(4) that a particular class of  CW solutions is the one given by
\begin {equation}\normalsize
U_{0}^{(1)}=i\sqrt{P^{(1)}} , \:\: U_{0}^{(2)}=i\sqrt{\alpha-P^{(1)}}
 \tag{8}
 \:,
 \end {equation}
with $P^{(2)}=\alpha-P^{(1)}$ and $g_{\rm s}=g_{\rm c}=1$, where $ 0 \leqslant P^{(1)}\leqslant \alpha$. The dispersion relation is given by
\begin {equation}\normalsize
\begin {split}
i\Omega_{(\pm,\pm)}=1\pm \left\{{-\frac{K^{4}}{4}+\frac{\alpha K^{2}}{2}[-\rm{sgn}(\bar\beta_2)\pm 1]}\right\}^{1/2}
\end {split}
 \tag{\normalsize{9}}
 \:.
 \end {equation}
Equation (9) implies that the MI only occurs for this particular class of solutions in the anomalous dispersion regime, for which  $\rm{sgn}(\bar\beta_2)=-1$, and only for $\alpha>1$ and $i\Omega=i\Omega_{(-,+)}$. 
The range of allowed wavenumbers at which the MI occurs is $K_-\le |K|\le K_+$, 
where $K_{\pm}=\left[{2\alpha\pm2\sqrt{\alpha^2-1}}\right]^{1/2}$. \\
\indent To verify the results of our analytical study for this particular case, we have performed a numerical integration of Eq.~(4),
using a symmetrized fast Fourier transform, split-step algorithm \cite{Fleck} with the following initial conditions
\begin {equation}\normalsize
U^{(j)}(\sigma,\tau=0)=U_{0}^{(j)}+{\rm{Re}}\left[x^{(j)}\exp\left(i K\sigma\right)\right]
 \notag
 \end {equation} 
\vspace{-6mm}
\begin{equation}
\hspace{2cm}+i{\rm{Re}}\left[y^{(j)}\exp\left(i K\sigma\right)\right],
\tag{10}
\end {equation}
with $j=1,2$, where $U_{0}^{(j)}$ is given by Eq.~(8) and $[x^{(1)},y^{(1)},x^{(2)},y^{(2)}]^T$ is the eigenvector corresponding to the eigenvalue $i\Omega_{(-,+)}$ calculated for the conditions that allow the MI to occur. Explicit expressions for the eigenvector components are
\begin {equation}
x^{(1)}=C,
\tag{11.a}
\end {equation}
\vspace{-6mm}
\begin {equation}
y^{(1)}=C\rm{sgn}(\bar\beta_2){K}^2/\Delta 
\tag{11.b}
\end {equation}
\vspace{-6mm}
\begin {equation}
x^{(2)}=-\frac{C{\rm{sgn}}(\bar\beta_2){K}^2[P^{(1)}(\alpha-P^{(1)})]^{1/2}}{\Delta^2/4+K^4/4+{\rm{sgn}(\bar\beta_2)}K^2(\alpha-P^{(1)})},
\tag{11.c}
\end {equation}
\vspace{-4mm}
\begin {equation}
y^{(2)}=-\frac{{C{K}^4[P^{(1)}(\alpha-P^{(1)})]^{1/2}}/\Delta}{\Delta^2/4+K^4/4+{\rm{sgn}(\bar\beta_2)}K^2(\alpha-P^{(1)})},
\tag{11.d}
\end {equation}
where $\Delta=2[1-i\Omega_{(-,+)}]$ and $C$ is an arbitrary constant that quantifies the modulation around the CW solutions. Here, we take $C=0.1$. 
The initial conditions described in Eq.~(10) and (11) are the CW solutions modulated by the solutions of the linearized system, {\it{i.e.}} Eq.~(5). In the spirit of the induced MI \cite{Hasegawa}, we expect that these initial conditions will initiate the formation of  trains of bright solitons. For the numerical integration, we use the following parameters: $\bar\beta_2=-0.01$, $\alpha=2$, $P^{(1)}=0.5$, $P^{(2)}=1.5$, $g_{\textrm{c}}=g_{\textrm{s}}=1$. For $\alpha=2$, the range of the allowed wavenumbers for the MI is $\left[{4-2\sqrt{3}}\right]^{1/2}\le |K|\le \left[{4+2\sqrt{3}}\right]^{1/2}$. In particular, we take $K=15(|\bar\beta_2|)^{1/2}$ for the case described in Fig.~1 and $K=8(|\bar\beta_2|)^{1/2}$ for the case described in Fig.~2. Figure 1 shows the formation of two trains of 15 mode-locked, bright solitons for $K=15(|\bar\beta_2|)^{1/2}$. In this case, the soliton trains have the same period as the initial conditions. On the contrary, in Fig.~2 we show soliton trains  for $K=8(|\bar\beta_2|)^{1/2}$,  where the period of the soliton train is twice that of the initial conditions. In both figures, we also show the Fourier transform of $|U^{(j)}|^2$, ${\mathcal{F}}_{t}[|U^{(j)}|^2|]$, which corresponds to the radio frequency spectrum that would be obtained after a photodetector, and which we have normalized with respect to the largest spectral component in both modes.\\
 \begin {figure}
\begin {center}
\includegraphics [scale=0.44] {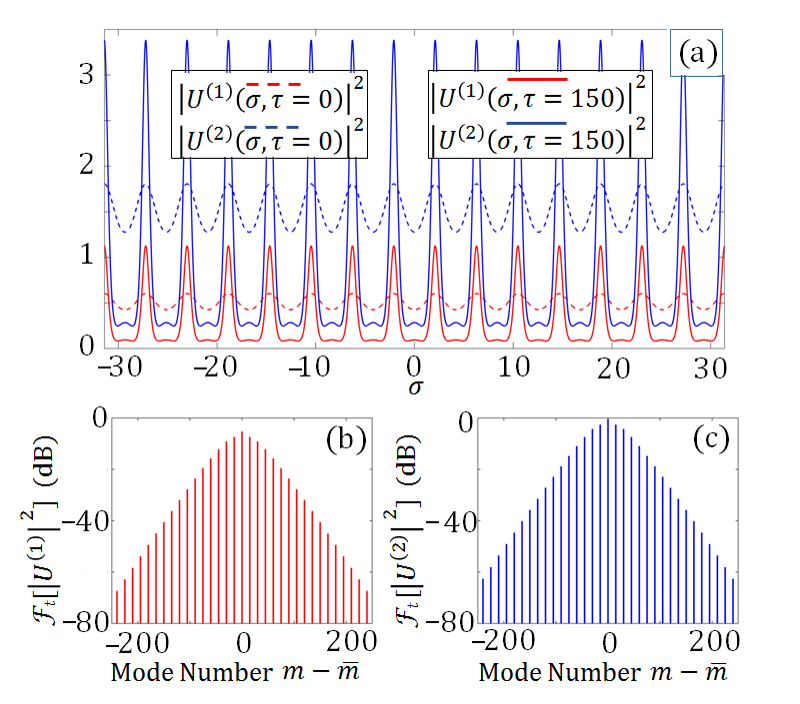}
\end {center}
\caption {(a): Two trains of 15 mode-locked bright solitons at $\tau=150$ calculated by the numerical integration of Eq.~(4) with the following parameters: $\bar\beta_2=-0.01$, $\alpha=2$, $P^{(1)}=0.5$, $P^{(2)}=1.5$, $g_{\textrm{c}}=g_{\textrm{s}}=1$. The input conditions (dashed curves) are described by Eq.~(10) and (11) with a wavenumber $K=15(|\bar\beta_2|)^{1/2}$. The lower curves refer to mode-1 and the upper curves refer to mode-2. The Fourier transform (${\mathcal{F}}_{t}$) of the intensity of the soliton trains (${\mathcal{F}}_{t}[|U^{(j)}|^2]$) is shown respectively in (b) for mode-1 and (c) for mode-2. The mode number spacing between two adjacent spectral lines is 15. 
}
\end {figure}
\begin {figure}
\begin {center}
\includegraphics [scale=0.44] {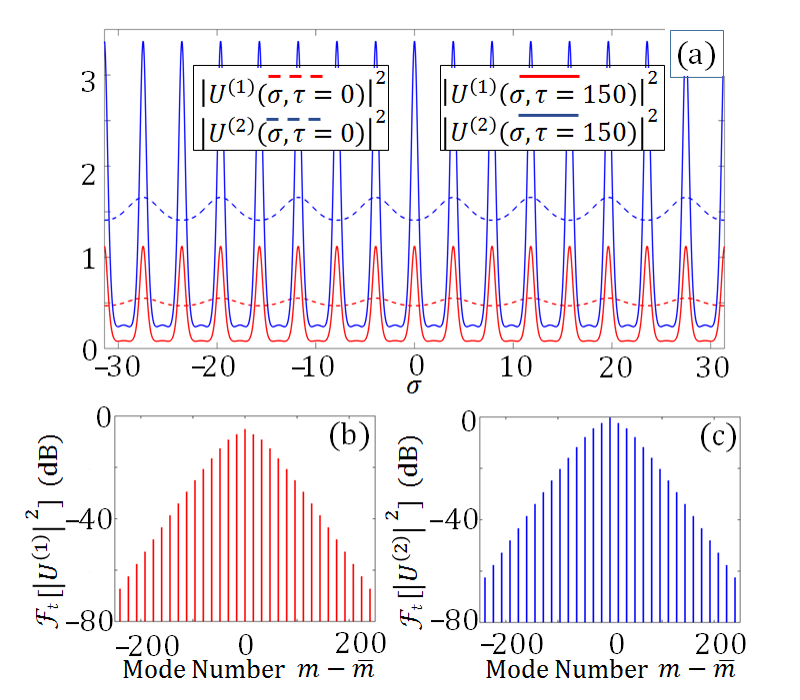}
\end {center}
\caption {Same parameters as in Fig.~1 except that here $K=8(|\bar\beta_2|)^{1/2}$. In (a) the lower curves refer to mode-1 and the upper curves refer to mode-2. The period of the soliton trains is twice that of the input conditions. In this case, the mode number spacing between two adjacent spectral lines is 16.}
\end {figure}
It is often assumed that the MI can only exist in the anomalous dispersion regime, as is the case for the solution that we just presented; however, we will show that the MI instability and bright solitons can also exist in the normal dispersion regime for Eq.~(4).  In fact, another class of CW solutions of Eq.~(4) is given by
\begin {equation}\normalsize
U_{0}^{(1)}=U_{0}^{(2)}=i\sqrt{\alpha/(1+g_{\rm c})} 
 \tag{12}
 \:,
 \end {equation}
where $P^{(1)}=P^{(2)}=\alpha/(1+g_{\rm c})$ , $g_{\rm s}=1$ and $\alpha>0$. Both the pump detuning and the cross coupling coefficient play a critical role. In this case, the dispersion is given by
\begin {equation}\normalsize
\begin {split}
i\Omega_{(\pm,\pm)}=1 \pm \left\{-\frac{K^{4}}{4}+\frac{\alpha K^{2}}{1+g_{\rm c}}[-\rm{sgn}(\bar\beta_2)\pm \it{g}_{\rm c}]\right\}^{1/2}
\end {split}
 \tag{13}
 \:.
 \end {equation}
In this case, the MI takes place both in the normal and in the anomalous dispersion regime. In particular, we find an instability in the normal dispersion regime when
\begin {equation}\normalsize
g_{\rm c}>1 ,\, \alpha>\frac{1+g_{\rm c}}{g_{\rm c}-1}, \ \textrm{and}\ i\Omega=i\Omega_{(-,+)}
 \tag{14}
 \:.
 \end {equation} 
The range of allowed wavenumbers in the MI region is: $K_-\le |K|\le K_+$, 
where
\begin {equation}
 K_{\pm}=\left\{\frac{2\alpha(g_c-1)}{g_c+1}\pm2\left[\frac{\alpha^2(g_c-1)^2}{(g_c+1)^2}-1\right]^{1/2}\ \right\}^{1/2}.
\tag{15}
\end {equation}
 We have performed a  large scale numerical integration of Eq.~(4) in the parameter space $(g_{\rm{c}},\alpha )$ to search for coupled bright solitons in the normal dispersion regime. We have used the initial conditions described in Eq.~(10), where, in this case,  $U_{0}^{(j)}$ is the CW solution given by Eq.~(12) and $[x^{(1)},y^{(1)},x^{(2)},y^{(2)}]^T$ is the eigenvector corresponding to the eigenvalue $i\Omega_{(-,+)}$. Explicit expressions  for the eigenvector components are
\begin {equation}
x^{(1)}=C,
\tag{16.a}
\end {equation}
\vspace{-6mm}
\begin {equation}
y^{(1)}=C\rm{sgn}(\bar\beta_2){K}^2/\Delta 
\tag{16.b}
\end {equation}
\vspace{-6mm}
\begin {equation}
x^{(2)}=-\frac{C{\rm{sgn}}(\bar\beta_2){K}^2 \alpha g_{\rm{c}}}{\Delta^2/4+K^4/4+{\rm{sgn}(\bar\beta_2)}K^2\alpha/(1+g_{\rm{c}})},
\tag{16.c}
\end {equation}
\vspace{-4mm}
\begin {equation}
y^{(2)}=-\frac{{C{K}^4\alpha g_{\rm{c}}}/[\Delta(1+g_{\rm{c}})]}{\Delta^2/4+K^4/4+{\rm{sgn}(\bar\beta_2)}K^2\alpha/(1+g_{\rm{c}})},
\tag{16.d}
\end {equation}
where $\Delta=2[1-i\Omega_{(-,+)}]$ and $C=0.1$. The wavenumber $K$ has been chosen near the center of the allowed range of wavenumbers in the MI region, \textit {i.e.} $K=(|\bar\beta_2|)^{1/2}p$ and $p=\left \lfloor{(K_{+}+K_{-})/(2(|\bar\beta_2|)^{1/2})}\right \rfloor$, where $\left \lfloor{\cdot}\right \rfloor$  is the floor function. The results of this computational  study are summarized in Fig.~3. Trains of mode-locked bright solitons are found near the border of the MI region. Figure 4 shows one example. 
\begin {figure}
\begin {center}
\includegraphics [scale=0.42] {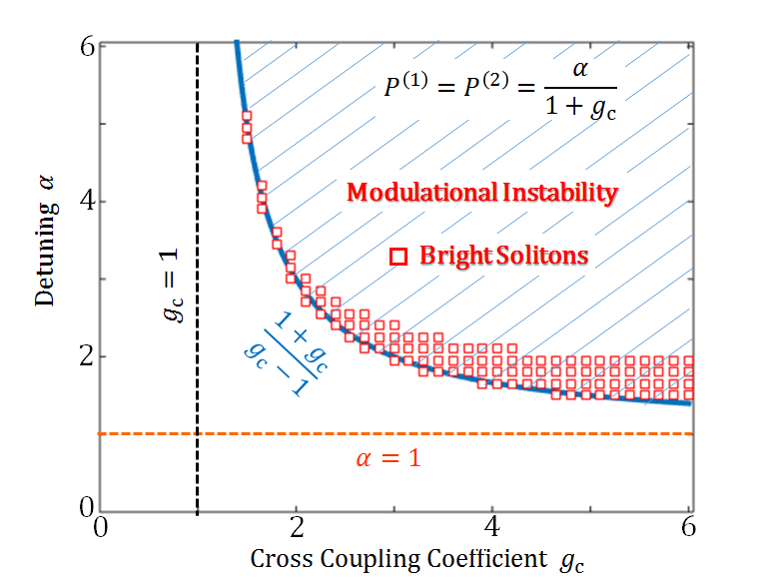}
\end {center}
\caption {Parameter values (shaded) in the parameter space with normal dispersion for which the MI occurs. The open squares indicate the position in the parameter space of the  coupled bright solitons  obtained by the numerical integration of Eq.(4) with $\bar\beta_2=0.01$. It is noted that bright solitons exist near the border of the MI region.}
\end {figure}
\begin {figure}
\begin {center}
\includegraphics [scale=0.45] {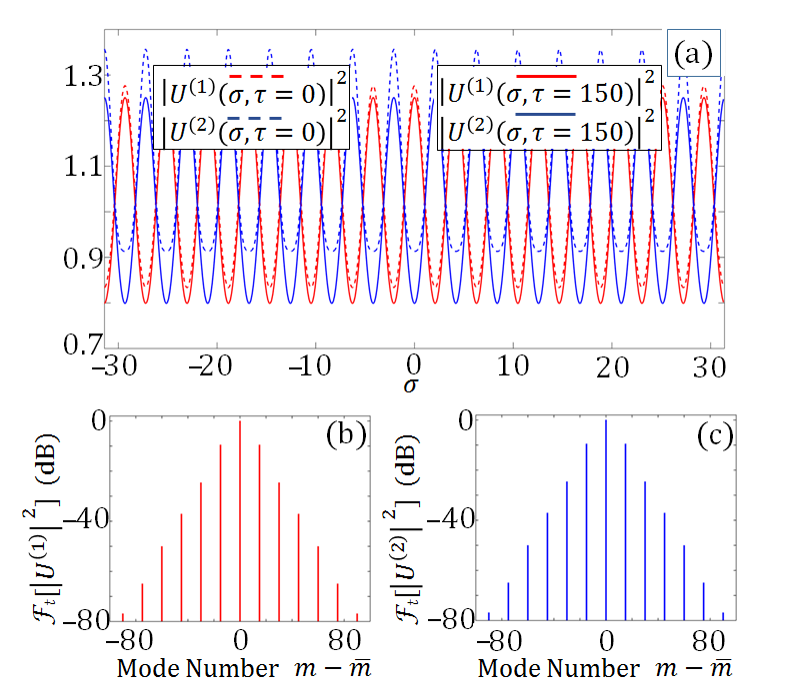}
\end {center}
\caption {(a): Two trains of 15 mode-locked bright solitons at $\tau=150$ calculated by the numerical integration of Eq.~(4) with the following parameters: $\bar\beta_2=0.01$, $\alpha=3.1$, $g_{\textrm{c}}=2$. The input conditions (dashed curves) are described by Eq.~(10) and (16)  with a wavenumber $K=15(|\bar\beta_2|)^{1/2}$. The lower dashed curve refers to mode-1 and the upper dashed curve refers to mode-2. The Fourier transform (${\mathcal{F}}_{t}$) of the intensity of the soliton trains (${\mathcal{F}}_{t}[|U^{(j)}|^2]$) is shown respectively in (b) for mode-1 and (c) for mode-2. The mode number spacing between two adjacent spectral lines is 15.}
\end {figure}

The existence of coupled bright solitons in the normal dispersion regime is  an important result from both a conceptual standpoint and  for possible applications. 
The MI in the normal dispersion regime and  new  solitary waves for the  standard coupled NLSE for optical fibers, {\it{i.e.}}, with no pump term, no detuning and no loss, has been studied in the past \cite {Haelterman}. However, in the context of WGM resonators, where the pump term, the detuning and the loss are all present and all play a fundamental role,  the existence of coupled bright solitons in the normal dispersion regime has never been predicted or studied before. This finding is potentially important for applications in light of the recent theoretical and experimental efforts aimed at obtaining bright solitons in normally dispersive WGM resonators \cite{Lobanov,Brasch}.
Currently fabricated WGM resonators for nonlinear frequency comb generation are based on dielectric materials, such as glass, which have anomalous dispersion in the near-IR and longer wavelengths. On the other hand, it would be highly desirable for many applications to achieve nonlinear frequency comb generation in the visible and near-UV range.  One way to obtain anomalous dispersion at shorter wavelengths is to counteract the natural dispersion of the material with the geometrical dispersion that is induced by modifying the resonator shape \cite{Li} and/or its boundary conditions \cite{Trillo,Coen1}. However, it is difficult  to obtain anomalous dispersion deep into the visible wavelengths. Other approaches recently proposed include the use of a phase/amplitude modulated pump instead of a pump at a fixed frequency \cite {Lobanov} or the use of soliton Cherenkov  radiation \cite {Brasch}.
The nonlinear mode-coupling that we have studied in this work may offer an alternative and more efficient approach for the generation of mode-locked trains of bright solitons in the normal dispersion regime.
\section {Conclusions}
\indent 
 In conclusion, we have studied nonlinear  mode coupling in WGM resonators and we have demonstrated the possibility of generating trains of mode-locked bright solitons in the normal dispersion regime.\\
\indent Recent experiments show that strong modification of the effective dispersion properties of the resonator with respect to the material properties can occur in spectral regions near the avoided-mode-crossing points of the resonator \cite {Gaeta,Weiner2}. In these regions, two frequency-degenerate guided modes of the resonator undergo a strong linear interaction with a GVM  practically equal to zero. This strong interaction leads to the formation of two new hybrid guided modes, that are no longer frequency-degenerate, whose frequency splitting depends on the  coupling strength of the frequency-degenerate modes.  This phenomenon may cause, for example, one of the two hybrid modes to  acquire anomalous GVD  in a spectral region that would otherwise be characterized by normal GVD. In such a scenario, it is important to study the nonlinear interaction of these hybrid modes, and Eq.~(3) can be used for this purpose by considering a null GVM ($\delta^{(j)}=0$). This topic will be the subject of future investigations.\\
\section {Acknowledgments}
We acknowledge financial support from  AMRDEC/DARPA project W31P4Q-14-1-0002  and ARL project W911NF-13-2-0010. The numerical simulations
were carried out at UMBC's high performance computing facility. We thank Andrew Weiner, Minghao Qi, and members of their research groups for suggesting this problem to us and for useful discussions.
\section {Appendix}
\indent Our starting point is the wave equation for the real electric field $\bar{\textbf{E}}$: 
\begin {equation}\normalsize
-\nabla\times\nabla \times\bar{\textbf{E}}-\frac{1}{c^2}\frac{\partial^2}{\partial t^2}\Bigg[\int\limits_{-\infty}^{+\infty}\hat\epsilon(\tau)\bar{\textbf{E}}(\textbf{r},t-\tau)d\tau+
\notag
\end {equation}
\vspace {-6mm}
\begin {equation}
\bar{\chi}^{(3)}(\bar{\textbf{E}}\cdot\bar{\textbf{E}})\bar{\textbf{E}}\Bigg]=-\frac{\omega_{p}^2\epsilon_{r,p}}{c^2}\bar{\textbf{E}}_{p}\cos(\omega_{p} t) \:,
\tag {A.1}
\end{equation}
where
\begin {equation}
 \hat\epsilon(\tau)=(1/2\pi)\int\limits_{-\infty}^{+\infty}\epsilon_{r}(\omega)\exp [-i\omega \tau]d\omega 
\tag {A.2}
\end {equation}
is the  linear dielectric response of the medium in the time domain with $\hat\epsilon(\tau<0)=0$ due to causality, $\epsilon_{r}(\omega)$ is the relative electric permittivity in the frequency domain, $\bar{\chi}^{(3)}$ is the cubic nonlinearity, $\bar{\textbf{E}}_p $ is the fraction of the pump field  coupled with the resonator and acts as the source term, $\omega_{p}$ the pump frequency, $\epsilon_{r,p}$ is the electric permittivity at the pump frequency, and $c$ the speed of light in the vacuum. Eq.~(A.1) is valid if we assume that the linear and nonlinear response of the material is local and isotropic and we also assume that the nonlinear response of the material is  instantaneous. The assumption that the response is instantaneous corresponds physically to just considering the fast nonlinear electronic response of the medium and neglecting the contribution of the molecular vibrations (Raman effect) \cite{Agrawal,Hasegawa1}.  It is convenient to pass from the real to the complex field representation: $\bar{\textbf{E}}=(1/2)[\textbf{E}+\textrm{complex conjugate}]$ and $\bar{\textbf{E}}_{p}\cos(\omega_{p} t)=(1/2)[\textbf{E}_{p}\exp(-i\omega_{p} t)+\textrm{complex conjugate}]$. With reference to Fig.~5, we can  expand the complex electric field \textbf{E} by using the guided modes of the resonator in cylindrical coordinates as
\begin {equation}\normalsize
\textbf{E}(\rho,\theta,z,t)=\sum\limits_{j,m} \ A_m^{(j)}(t)\textbf{F}_m^{(j)}(\rho,z) \exp{ [i(m\theta-\omega_m^{(j)} t)]}\:,
\tag {A.3}
\end {equation}
and we can expand  $\textbf{E}_p $  as 
\begin {equation}\normalsize
\textbf{E}_p(\theta)= \textbf{e}_p\sum\limits_{m} E_{p,m}\exp{(i m\theta)}\:,
\tag {A.4}
\end{equation}
where the $A_m^{(j)}(t)$ are the time-dependent envelope functions, $j=1,2$ labels the two families of transverse modes  with amplitudes $\textbf{F}_m^{(1)}(\rho,z)$ and $\textbf{F}_m^{(2)}(\rho,z)$, and eigenfrequencies $\omega_m^{(1)}$ and $\omega_m^{(2)}$, $\textbf{e}_p$ is the polarization vector of the pump field, $\theta$ is the azimuthal angle and finally $m\in[1,2,...,N]$ is the azimuthal number that labels the eigenfrequencies in each family. We note that, while  we restrict our analysis here to two families of modes, our coupled mode theory can be extended to an arbitrary number of families.
\begin {figure}
\begin{center}
\includegraphics [scale=0.40] {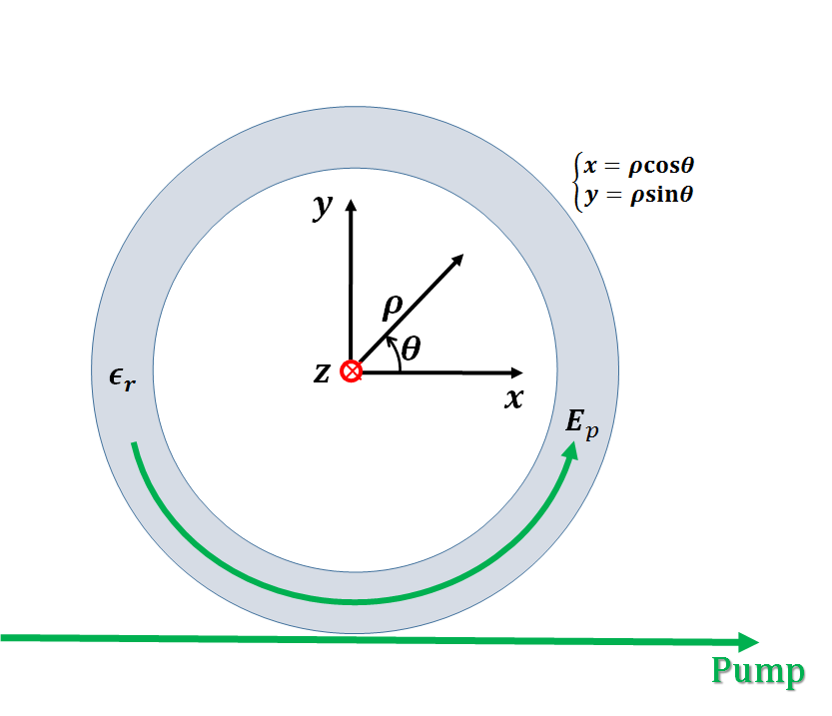}
\end{center}
\caption {Sketch of the geometry investigated. A pump field is coupled to an axially symmetric  WGM resonator placed in the vacuum. The resonator has an azimuthal angle $\theta$ and relative electric permittivity $\epsilon_r$. The field $\textbf{E}_p$ is the fraction of the  pump field that is coupled to the resonator.}
\end {figure}
The field profile $\tilde{\textbf{F}}_m^{(j)}(\rho,\theta,z)={\textbf{ F}}_m^{(j)}(\rho,z)\exp(im\theta)$ solves the eigenmode equation
\begin {equation}\normalsize
\nabla\times\nabla \times\tilde{\textbf{F}}_m^{(j)}(\rho,\theta,z)=\frac{\epsilon_{r}(\omega_m^{(j)})\omega_m^{(j)2}}{c^2}\tilde{\textbf{F}}_m^{(j)}(\rho,\theta,z)\:,
\tag {A.5}
\end{equation}
subject to the  orthonormality condition
\begin {equation}
 (1/V_{c})\int_{V}\tilde{\textbf{F}}_m^{(j)}\cdot\tilde{\textbf{F}}_{m'}^{(j')*}dV =\delta_{m,m'}\delta_{j,j'}\:, 
\tag{A.6}
\end {equation}
where $dV=\rho d\rho d\theta dz$ is the elementary volume in cylindrical coordinates, $V_c$ is the volume occupied by the resonator, and $\delta_{k,l}$ is the Kr\"onecker $\delta$-function.  In Eq.~(A.5) we take into account the material as well as the waveguide dispersion, since we explicitly consider the electric permittivity as a function of frequency.
\indent Expressing Eq.~(A.1) in the complex field representation and using Eqs.~(A.3) and (A.4), as well as the eigenmode equation, Eq.~(A.5), and, finally, by invoking the slowly varying envelope approximation $|\ddot A_m^{(j)}|\ll\omega_m^{(j)}|\dot A_m^{(j)}|\ll\omega_m^{(j)2} |A_m^{(j)}|$, the dot and double dot denote respectively the first and second time derivative, we arrive at the following equation, containing just the first order time derivatives of the envelope functions, 
\begin {equation}\normalsize
\sum\limits_{j,m}\epsilon_{r}(\omega_m^{(j)})\omega_m^{(j)} \dot A_m^{(j)}(t) \textbf{F}_m^{(j)} \exp{[i (m\theta-\omega_m^{(j)} t)]}=
\notag
\end {equation}
\vspace {-4mm}
\begin {equation}\normalsize
i{\chi^{(3)}}\sum\limits_{j,m}\omega_m^{(j)2} A_m^{(j)}(t)|\textbf{E}|^2 \textbf{F}_m^{(j)} \exp{[i (m\theta-\omega_m^{(j)} t)]}  
\notag
\end{equation}
\vspace {-4mm}
\begin {equation}\normalsize
+i\frac{\chi^{(3)}}{2}\sum\limits_{j,m}\omega_m^{(j)2} A_m^{(j)*}(t)(\textbf{E}\cdot\textbf{E})\textbf{F}_m^{(j)*} \exp{[-i (m\theta-\omega_m^{(j)} t)]}
\notag
\end{equation}
\vspace {-4mm}
\begin {equation}
+\frac{1}{2}i\omega_p^2\epsilon_{r,p}\exp{(-i\omega_p t)}
\textbf{e}_p\sum\limits_{m}{E_{p,m}\exp{(i m\theta)}},
\tag {A.7}
\end{equation}
where $\chi^{(3)}=\bar{\chi}^{(3)}/4$ and
\begin {equation}
 |\textbf{E}|^2=\sum\limits_{\alpha,\beta,\gamma,\delta}A_\alpha^{(\gamma)}A_\beta^{(\delta)*}\textbf{F}_\alpha^{(\gamma)}\cdot\textbf{F}_\beta^{(\delta)*}\times
\notag 
\end {equation}
\vspace {-6mm}
\begin {equation}
\exp\{i [(\alpha-\beta)\theta-(\omega_\alpha^{(\gamma)}-\omega_\beta^{(\delta)}) t]\}\:,
\tag {A.8}
\end {equation}

\begin {equation}
\textbf{E}\cdot\textbf{E}=\sum\limits_{\alpha,\beta,\gamma,\delta} A_\alpha^{(\gamma)}A_\beta^{(\delta)}\textbf{F}_\alpha^{(\gamma)}\cdot\textbf{F}_\beta^{(\delta)}\times
\notag 
\end {equation}
\vspace {-6mm}
\begin {equation}
\exp{\{i [(\alpha+\beta)\theta-(\omega_\alpha^{(\gamma)}+\omega_\beta^{(\delta)}) t]}\}\:,
\tag {A.9}
\end {equation}
with $(\alpha,\beta)\in[1,2,...,N]$ and $(\gamma,\delta)\in[1,2]$. From now on, we omit the dependence of the envelope functions on $t$ and the dependence of the field profiles on $\rho$, $\theta$ and $z$.
In order to arrive at coupled mode equations, we project Eq.~(A.7) on the  modes, and we use their orthonormality, which yields 
\begin {equation}\normalsize
\dot A_\eta^{(1)} =
i\frac{\chi^{(3)}}{\epsilon_{r}(\omega_\eta^{(1)})\omega_\eta^{(1)}}\sum\limits_{\alpha,\beta,\gamma,\delta,j} A_{\eta-\alpha+\beta}^{(j)}A_{\alpha}^{(\gamma)}A_{\beta}^{(\delta)*}\omega_{\eta-\alpha+\beta}^{(j)2}\times
\notag
\end {equation}
\vspace{-4mm}
\begin {equation}
 \exp{[i (\omega_\eta^{(1)}-\omega_{\eta-\alpha+\beta}^{(j)}-\omega_\alpha^{(\gamma)}+\omega_\beta^{(\delta)}) t]}G_{\eta,\alpha,\beta}^{(j,1)(\gamma,\delta)}
\notag
\end {equation}
\vspace{-4mm}
\begin {equation}
+i\frac{\chi^{(3)}}{2\epsilon_{r}(\omega_\eta^{(1)})\omega_\eta^{(1)}}\sum\limits_{\alpha,\beta,\gamma,\delta,j} A_{-\eta+\alpha+\beta}^{(j)*}A_{\alpha}^{(\gamma)}A_{\beta}^{(\delta)}\omega_{-\eta+\alpha+\beta}^{(j)2} \times
\notag
\end {equation}
\vspace{-4mm}
\begin {equation}
\exp{[i (\omega_\eta^{(1)}+\omega_{-\eta+\alpha+\beta}^{(j)}-\omega_\alpha^{(\gamma)}-\omega_\beta^{(\delta)} )t]}H_{\eta,\alpha,\beta}^{(j,1)(\gamma,\delta)}
\notag
\end {equation}
\vspace{-4mm}
\begin {equation}
-\frac{A_\eta^{(1)}}{\tau_{\eta}^{(1)}}+\frac{1}{2}i\omega_p^2\frac{\epsilon_{r,p}}{\epsilon_{r}(\omega_\eta^{(1)})\omega_\eta^{(1)}}\exp[i(\omega_\eta^{(1)}-\omega_p )t]P_\eta^{(1)},
\tag{A.10.a}
\end{equation}
\begin {equation}\normalsize
\dot A_\eta^{(2)} =
i\frac{\chi^{(3)}}{\epsilon_{r}(\omega_\eta^{(2)})\omega_\eta^{(2)}}\sum\limits_{\alpha,\beta,\gamma,\delta,j} A_{\eta-\alpha+\beta}^{(j)}A_{\alpha}^{(\gamma)}A_{\beta}^{(\delta)*}\omega_{\eta-\alpha+\beta}^{(j)2}\times
\notag
\end {equation}
\vspace{-4mm}
\begin {equation}
 \exp[i (\omega_\eta^{(2)}-\omega_{\eta-\alpha+\beta}^{(j)}-\omega_\alpha^{(\gamma)}+\omega_\beta^{(\delta)}) t]G_{\eta,\alpha,\beta}^{(j,2)(\gamma,\delta)}
\notag
\end {equation}
\vspace{-4mm}
\begin {equation}
+i\frac{\chi^{(3)}}{2\epsilon_{r}(\omega_\eta^{(2)})\omega_\eta^{(2)}}\sum\limits_{\alpha,\beta,\gamma,\delta,j} A_{-\eta+\alpha+\beta}^{(j)*}A_{\alpha}^{(\gamma)}A_{\beta}^{(\delta)}\omega_{-\eta+\alpha+\beta}^{(j)2}\times
\notag
\end {equation}
\vspace{-4mm}
\begin {equation}
 \exp[i (\omega_\eta^{(2)}+\omega_{-\eta+\alpha+\beta}^{(j)}-\omega_\alpha^{(\gamma)}-\omega_\beta^{(\delta)}) t]H_{\eta,\alpha,\beta}^{(j,2)(\gamma,\delta)}
\notag
\end {equation}
\vspace{-4mm}
\begin {equation}
-\frac{A_\eta^{(2)}}{\tau_{\eta}^{(2)}}+\frac{1}{2}i\omega_p^2\frac{\epsilon_{r,p}}{\epsilon_{r}(\omega_\eta^{(2)})\omega_\eta^{(2)}}e^{i[\omega_\eta^{(2)}-\omega_p ]t}P_\eta^{(2)}.
\tag{A.10.b}
\end{equation}
We have introduced the following overlap integrals, involving just the transverse profile of the guided modes in the resonator $\textbf{F}_m^{(j)}(\rho,z)$,
\begin {equation}\normalsize
P_\eta^{(j)}=\frac{E_{p,\eta}}{V_c}\int\limits_{V_c}\textbf{F}_{\eta}^{(j)*}\cdot\textbf{e}_{p}\,dV \: ,\tag{A.11.a}
\end {equation}
\begin {equation}\normalsize
G_{\eta,\alpha,\beta}^{(j,j')(\gamma,\delta)}=\frac{1}{V_c}\int \limits_{V_c}(\textbf{F}_{\eta-\alpha+\beta}^{(j)}\cdot\textbf{F}_{\eta}^{(j')*})(\textbf{F}_\alpha^{(\gamma)}\cdot\textbf{F}_\beta^{(\delta)*})\,dV \: ,\tag{A.11.b}
\end {equation}
\begin {equation}\normalsize
H_{\eta,\alpha,\beta}^{(j,j')(\gamma,\delta)}=\frac{1}{V_c}\int \limits_{V_c}(\textbf{F}_{-\eta+\alpha+\beta}^{(j)*}\cdot\textbf{F}_{\eta}^{(j')*})(\textbf{F}_\alpha^{(\gamma)}\cdot\textbf{F}_\beta^{(\delta)})\,dV \: .\tag{A.11.c}
\end {equation}
The integration for the overlap integrals is extended only over the resonator volume $V_c$ because $\chi^{(3)}$ is zero outside the resonator, and $\textbf{E}_p$ is by definition the fraction of the pump field coupled with the resonator. 
The coefficient $P_\eta^{(j)}$ is the effective field for the mode $\eta$ of the family $j$, while the terms  $\chi^{(3)}G_{\eta,\alpha,\beta}^{(j,j')(\gamma,\delta)}$ and $\chi^{(3)}H_{\eta,\alpha,\beta}^{(j,j')(\gamma,\delta)}$ are the effective nonlinear coupling coefficients, due respectively to the self-phase modulation (SPM) and the four-wave mixing (FWM) cubic nonlinearity. Equation (A.10) describes two nonlinear four-wave mixing processes. The first is due to the SPM cubic nonlinearity and has a frequency detuning given by $[\omega_\eta^{(1,2)}-\omega_{\eta-\alpha+\beta}^{(j)}-\omega_\alpha^{(\gamma)}+\omega_\beta^{(\delta)}] $ for the families 1 and 2. The second is due to the FWM cubic nonlinearity and has a frequency detuning given by $[\omega_\eta^{(1,2)}+\omega_{-\eta+\alpha+\beta}^{(j)}-\omega_\alpha^{(\gamma)}-\omega_\beta^{(\delta)}] $ for the families 1 and 2. The detuning for both processes would be zero if the eigenfrequencies were equidistant, which would correspond to perfect phase matched interactions and infinite coherence length. In practice, the perfect phase matching condition is never fulfilled in  WGM resonators where instead the deviation from equidistance of the eigenfrequencies  plays a fundamental role, as we will show later. The effect of the finite bandwidth of the cavity modes has been taken into account in Eq.~(A.10) by the phenomenological introduction of the decay terms $-A_\eta^{(j)}/\tau_m^{(j)}$ into the equations, where $\tau_m^{(j)}=2/\Delta\omega_m^{(j)}$ is the  photon lifetime in the cavity and $\Delta\omega_m^{(j)}$ is the bandwidth of the resonance.\\ 
\indent Equation (A.10) is an exact representation of the electromagnetic problem stated in Eq.~(A.1). However, the direct numerical integration of these coupled mode equations is computationally inefficient because it is necessary to integrate a system of $2N$ equations, each one containing $2\cdot 8\cdot N^2$ terms, and a microresonator typically contains $N\sim1000$ to $10,000$ modes.
We now decouple the transverse field evolution  from its azimuthal evolution by using only  two dominant modes for the transverse field profile, one for each family, namely:
$\textbf{F}_{\bar m}^{(j)} (\rho,z)$ with $j=1,2$ where  $\bar m$  is the azimuthal number corresponding to the closest to the pump eigenfrequency for each family, i.e $\omega_p\cong \omega_{\bar m}^{(j)}$ for $j=1,2$. This approximation is justified because the dependence of the transverse field profile of a guided mode on the propagation wavevector---in our case on the azimuthal number $m$---is usually weak. Hence, we may assume that the transverse profile is nearly the same as for the respective dominant modes. In this way, Eq.~(A.3) can be rewritten in the following form
\begin {equation}\normalsize
\textbf{E}(\rho,\theta,z,t)=\sum\limits_{j=1}^{2} \textbf {F}_{\bar m}^{(j)}(\rho,z) 
\exp[i ({\bar m}\theta-\omega_{\bar m}^{(j)} t)] \Psi^{(j)} (\theta,t) \:,
\tag{A.12.a}
\end{equation}
where
\begin {equation}\normalsize
\Psi^{(j)} (\theta,t)=\sum\limits_{m=1}^{N} \ A_{m}^{(j)}(t) \exp\{i [(m-{\bar m})\theta-(\omega_{m}^{(j)}-\omega_{\bar m}^{(j)}) t]\}  \:,
\tag{A.12.b}
\end{equation}
is the spatio-temporal  envelope of the total field. By taking the partial time derivative of Eq.~(A.12.b) we obtain:
\begin {equation}\normalsize
\frac{\partial \Psi^{(j)} (\theta,t)}{\partial t}=\sum\limits_{m=1}^{N}[\ \dot A_{m}^{(j)}(t)-i (\omega_{m}^{(j)}-\omega_{\bar m}^{(j)}) A_{m}^{(j)}(t)]\times
\notag
\end {equation}
\vspace{-4mm}
\begin {equation}
\exp\{i [(m-{\bar m})\theta-(\omega_{m}^{(j)}-\omega_{\bar m}^{(j)}) t]\}  \:,
\tag{A.13}
\end{equation}
where the term $(\omega_{m}^{(j)}-\omega_{\bar m}^{(j)})$ can be expressed through a Taylor expansion as
\begin {equation} 
\omega_{m}^{(j)}-\omega_{\bar m}^{(j)}=\sum\limits_{k=1}^{n}(\zeta_k^{(j)}/k!)(m-\bar m)^k \:.
\tag {A.14} 
\end {equation}
The coefficient  $\zeta_1^{(j)}=(1/2) (\omega_{\bar m+1}^{(j)}-\omega_{\bar m-1}^{(j)})$ is  the free spectral range (FSR) of the resonator at the frequency $\omega_{\bar m}^{(j)}$ and the coefficient $\zeta_2^{(j)}=\omega_{\bar m+1}^{(j)}-2\omega_{\bar m}^{(j)}+\omega_{\bar m-1}^{(j)}$ equals, at lowest order, the deviation from  equidistance of the eigenfrequencies adjacent to  $\omega_{\bar m}^{(j)}$ and plays a role analogous to the group velocity dispersion (GVD) of a standard optical fiber. In particular, $\zeta_2^{(j)}<0$ corresponds to the normal dispersion regime in which the group velocity decreases for increasing frequencies, while $\zeta_2^{(j)}>0$ corresponds to the anomalous dispersion. It is also customary to introduce the GVD parameter $\beta_2^{(j)}=-\zeta_2^{(j)}=2\omega_{\bar m}^{(j)}-\omega_{\bar m+1}^{(j)}-\omega_{\bar m-1}^{(j)}$, so that the dispersion is normal when $\beta_2^{(j)}>0$ and anomalous when $\beta_2^{(j)}<0$. Using a Taylor expansion of the term $(\omega_{m}^{(j)}-\omega_{\bar m}^{(j)})$ and the following identity
\begin {equation}
(-i)^k\frac{\partial^k \Psi^{(j)}}{\partial \theta^k}=\sum\limits_{m=1}^{N} (m-{\bar m})^k A_{m}^{(j)}\times
\notag
\end {equation}
\vspace {-6mm}
\begin {equation}
\exp\{i [(m-{\bar m})\theta-(\omega_{m}^{(j)}-\omega_{\bar m}^{(j)}) t]\}\:, 
\tag {A.15}
\end {equation}
we can recast Eq.~(A.13) in the form
\begin {equation}\normalsize
\frac{\partial \Psi^{(j)}}{\partial t}=\sum\limits_{k=1}^{n}\ (-i)^{k+1}\frac{\zeta_k^{(j)}}{k!}\frac{\partial^k \Psi^{(j)}}{\partial \theta^k}
\notag
\end {equation}
\vspace {-6mm}
\begin {equation}
+\sum\limits_{m=1}^{N} \dot A_{m}^{(j)}\exp\{i [(m-{\bar m})\theta-(\omega_{m}^{(j)}-\omega_{\bar m}^{(j)}) t]\}\:,
\tag{A.16}
\end{equation}
where the time derivatives of the field envelopes in the the last term at the right-hand side can be explicitly calculated by using the coupled mode Eq.~(A.10).
Our goal is to write Eq.~(A.16) in a form that just includes envelope fields $\Psi^{(j)}(\theta,t)$, so that we obtain coupled nonlinear wave equations for $\Psi^{(j)}(\theta,t)$. In doing so, we make several approximations. First, we suppose that the decay times for all the modes of the same family are the same: $\tau_{m}^{(j)}=1/\tau^{(j)}$.  For high-Q WGM resonators, we generally have decay times on the order of  $\tau^{(j)} \sim1~\mu \rm{s}$ and Q-factors given by $Q^{(j)}=\omega_m^{(j)}/\Delta\omega_m^{(j)}\sim10^9$.
Second, we suppose that, as is usual in nonlinear optical phenomena, the effects of the nonlinearity and of the pump  on the envelope field occur on a much slower time scale than the time scale necessary for the field to complete one round-trip in the resonator. In a typical WGM resonator with $\sim 1~\rm {mm}$ radius, the round trip time is $\sim100~\rm{ps}$, while the time scale on which the nonlinearity and the pump field produce significant effects on the field envelope is in the $\rm{\mu s}$ or $\rm{ms}$ range. Hence, once the time derivative of the field envelopes is calculated using Eq.~(A.10), the nonlinear terms and pump terms in Eq.~(A.16) can be averaged over the azimuthal coordinate $\theta$ from $0$ to $2\pi$. Hence, all the terms proportional to $\exp[i (m-{\bar m})\theta]$,  with $m\neq \bar m$, do not effectively contribute to the process because they average to zero and can be neglected. Third, we  approximate the overlap integrals as follows: $G_{\bar m,\alpha,\beta}^{(s,j)(\gamma,\delta)}\cong G_{\bar m,\bar m,\bar m}^{(s,j)(\gamma,\delta)}$ and $H_{\bar m,\alpha,\beta}^{(s,j)(\gamma,\delta)}\cong H_{\bar m,\bar m,\bar m}^{(s,j)(\gamma,\delta)}$, consistent with the weak dependence of the radial field profiles on the azimuthal number.
Moreover, in the pump term we simplify  $\epsilon_{r,p}/\epsilon_{r}(\omega_{\bar m}^{(j)})\cong 1$ and $\omega_{p}/\omega_{\bar m}^{(j)}\cong 1$, and we introduce the detuning of the pump field with respect to the dominant modes, $\delta\omega^{(j)}=\omega_{\bar m}^{(j)}-\omega_p$. Fourth, we expand $\omega_{\bar m-\alpha+\beta}^{(s)2}$ and $\omega_{-\bar m+\alpha+\beta}^{(s)2}$ around $m=\bar m$ and keep only the lowest order. Finally,  we collect the nonlinear terms oscillating with the same detuning and retain only  the nonlinear terms whose frequency detuning vanishes,{\it{ i.e.}}, we only retain  the frequency-matched terms $\omega_{\bar m }^{(j)}-\omega_{\bar m}^{(s)}-\omega_{\bar m}^{(\gamma)}+\omega_{\bar m}^{(\delta)}=0$. We then obtain two incoherently coupled, externally driven, damped, generalized NLSEs or LLEs 
\begin {equation}\normalsize
\frac{\partial \Psi^{(j)}}{\partial t}=\sum\limits_{k=1}^{n}\ (-i)^{k+1}\frac{\zeta_k^{(j)}}{k!}\frac{\partial^k \Psi^{(j)}}{\partial \theta^k}-\frac{1}{\tau^{(j)}}\Psi^{(j)}(\theta,t)
\notag
\end{equation}
\vspace {-8mm}
\begin {equation}
+i\frac{P_{\bar m}^{(j)}\omega_p}{2} \exp[i\delta\omega^{(j)}t]
\notag
\end{equation}
\vspace {-8mm}
\begin {equation}\normalsize
+i\chi^{(3)} \omega_{\bar m}^{(j)}\Psi^{(j)}\sum\limits_{l=1}^{2}D^{(j,l)}|\Psi^{(l)}|^2, 
 \tag{A.17}
 \end {equation}
 where
\begin {equation}
 D^{(j,l)}=\frac{2^{(|j-l|-1)}}{\epsilon_{r}(\omega_{\bar m}^{(j)})V_c}\times
\notag
\end {equation}
\vspace {-6mm}
\begin {equation}
\int \limits_{V_c}[|\textbf{F}_{\bar m}^{(j)}|^2|\textbf{F}_{\bar m}^{(l)}|^2+|\textbf{F}_{\bar m}^{(j)}\cdot\textbf{F}_{\bar m}^{(l)}|^2+|\textbf{F}_{\bar m}^{(j)}\cdot\textbf{F}_{\bar m}^{(l)*}|^2]\,dV\:,
\tag{A.18}
\end {equation} 
are the overlap integrals of the interacting modes. Note that $D^{(1,2)}=D^{(2,1)}$. Equations (A.17) and (A.18) are our starting points, Eqs.~(1) and (2).

\end{document}